\documentclass[fleqn,usenatbib]{mnras}

\usepackage{newtxtext,newtxmath}

\usepackage[T1]{fontenc}
\usepackage{ae,aecompl}

\usepackage{graphicx}	
\usepackage{amsmath}	
\usepackage{amssymb}	

\title[Homogeneous abundances among eP1 stars]{Constant light element abundances suggest that the extended P1 in NGC 2808 is not a consequence of CNO-cycle nucleosynthesis.}

\author[Cabrera-Ziri et al.]{
I. Cabrera-Ziri,$^{1}$\thanks{Hubble Fellow}\thanks{E-mail: ivan.cabrera@cfa.harvard.edu}
C. Lardo,$^{2}$
and A. Mucciarelli$^{3,4}$
\\
$^{1}$Harvard-Smithsonian Center for Astrophysics, 60 Garden Street, Cambridge, MA 02138, USA\\
$^{2}$Laboratoire d'Astrophysique, Ecole Polytechnique F\'ed\'erale de Lausanne, Observatoire de Sauverny, CH-1290 Versoix, CH\\
$^{3}$Dipartimento di Fisica e Astronomia, Universit\`a degli Studi di Bologna, via Piero Gobetti 93/2, 40129, Bologna, IT\\
$^{4}$INAF-Astrophysics and Space Science Observatory Bologna, via Piero Gobetti 93/3, 40129, Bologna, IT
}

\date{Accepted XXX. Received YYY; in original form ZZZ}

\pubyear{2019}

\begin{document}
\label{firstpage}
\pagerange{\pageref{firstpage}--\pageref{lastpage}}
\maketitle

\begin{abstract}
Recent photometric results have identified a new population among globular cluster stars. This population, referred to as the ``extended P1'', has been suggested to be the manifestation of a new abundance pattern where the initial mass fraction of He changes among cluster stars that share the same CNO values. The current paradigm for the formation of the multiple stellar populations in globular clusters assumes that variations in He are the product of chemical ``enrichment'' by the ashes of the CNO-cycle (which changes He and other elements like C, N and O simultaneously). We obtained MIKE@Magellan spectra of six giant stars in NGC 2808, a cluster with one of the strongest examples of the extended P1 population. We provide the first complete characterization of the light elements abundances  for the stars along a significant range of the extended P1 photometric group. The stars from our sample appear to be homogeneous in C, N, O, Na, Mg and Al. The lack of a significant change in these products of the CNO-cycle, suggest that unlike the rest of the populations identified to date, the photometric changes responsible for the extended P1 feature are a consequence of an alternative mechanism. Our measurements, are consistent with the interpretations where the  changes of the He mass fraction among these stars could be consequence of p-p chain nucleosynthesis (which could increase the He in stars without affecting heavier elements). Having said that, direct measurements of He are necessary to conclude if variations of this element are present among extended P1 stars.
\end{abstract}




\section{Introduction}

It is a well established fact that globular clusters host multiple stellar populations  with different abundance variations of specific light elements. These variations are not random but instead follow distinct correlation patterns between different elements \citep[cf.][for recent reviews]{Charbonnel16,BL18}. Evidence for the multiple stellar populations have been detected in clusters spanning a large range of metallicities, masses, ages and galactic environments \citep[e.g.][]{Larsen14,Niederhofer17,Milone17,Schiavon17b,Martocchia18}.

For decades it has been known that populations of stars with different light elements abundances occupate different regions of the colour-magnitude diagram in a given evolutionary stage \citep[e.g.][]{Bond69,Osborn76,Marino08}. Today we can identify distinct sub-populations with extreme efficacy when precise abundances and ultraviolet photometry are combined \citep[e.g.][]{Milone15,Carretta18}. 

\begin{figure}
\centering
    \includegraphics[width=6cm]{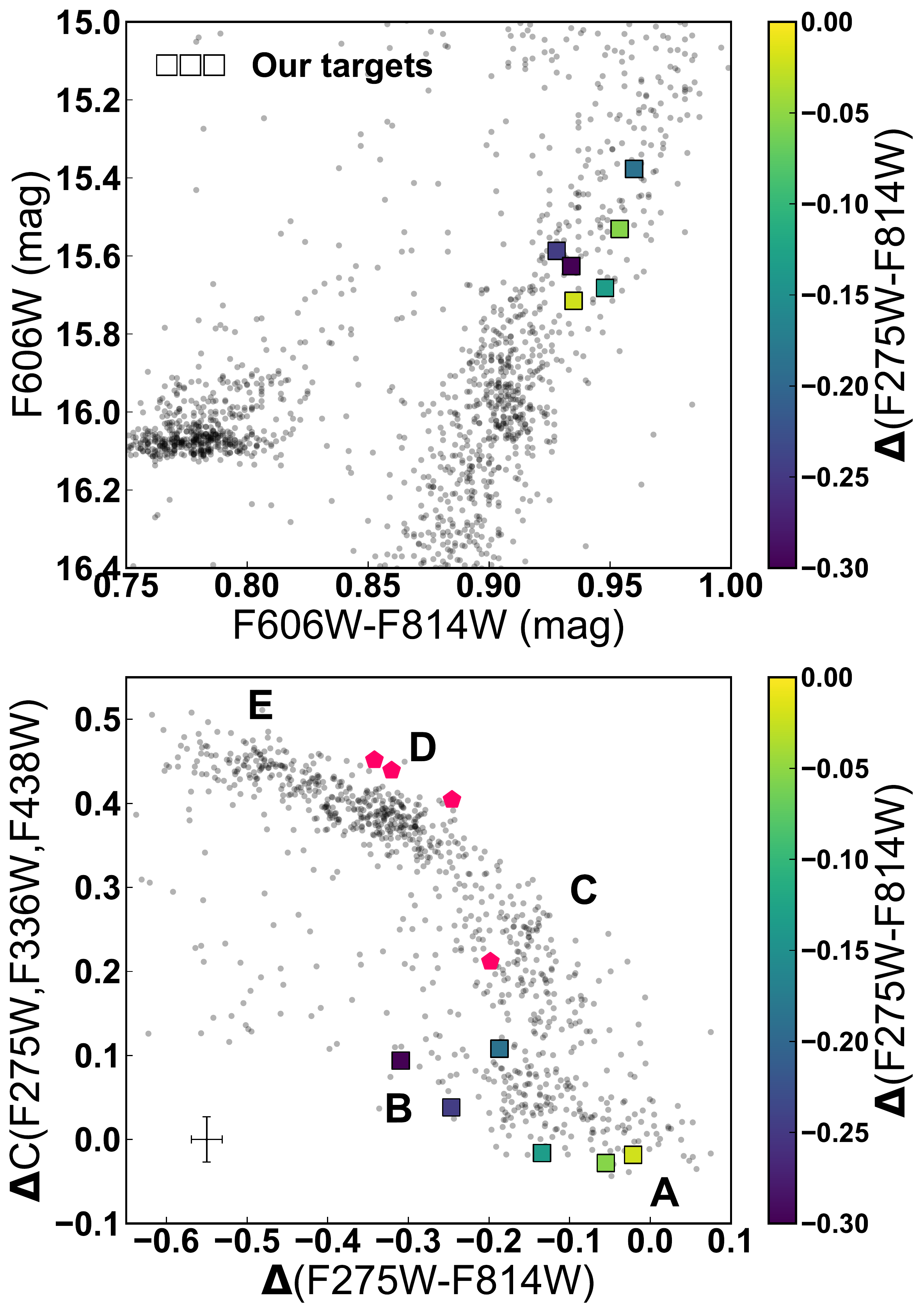}
    \caption{Squares represent the stars in our spectroscopic study while the small dots other NGC 2808 stars. \textbf{Upper panel:} position of our targets in the optical CMD. \textbf{Lower panel:} Chromosome map of NGC 2808 RGB stars. Colour coded is the position of the stars in our sample in the horizontal axis of the chromosome map. The letters show the general position of the five populations identified by \protect\cite{Milone15}: A and B with field like N (i.e. P1) while B, C and D with enhanced N with respect of field stars (i.e. P2). Our sample is distributed along the entire P1, sampling both population A and B. The pink pentagons show P2 stars with archival GIRAFFE, we show them here and in  Fig.~\ref{fig:abu} to illustrate how these populations map from  photometric to abundance space (no match found between UVES data and stars in the Lardo et al. chromosome map). The error bars represent the mean uncertainties of all the data points in the lower panel.}
    \label{fig:cmd_cmap}
\end{figure}

\cite{Milone15-M2} introduced a pseudo colour-colour plot (a.k.a. chromosome map) using F275W, F336W, F438W and F814W HST filters that allows a clear distinction between photometric sub-populations (e.g. bottom panel of Fig. \ref{fig:cmd_cmap}). These photometric sub-populations have different abundances of light elements like Na, Mg and Al \citep[e.g.][]{Carretta18}, however, these broad-band photometric filters are not sensitive to changes in the abundances of these particular elements but are sensitive to C, N, O and He instead \citep{Sbordone11}.

Recent work has been focused on the interpretation of the chromosome map, with the main goal of identifying what drives the position of the stars in this plane. The approach has been to compute stellar atmospheres with different stellar parameters and relative abundances between elements, then synthesise the photometry for each case and see where it lies in the chromosome map \citep[e.g.][]{Lardo18,Milone18}. There is a general consensus among these studies in that vertical changes in the chromosome map are mostly due to N variations while horizontal changes respond to variations in temperature, a consequence of different He mass fractions of the stars.\footnote{For completeness, it has been also pointed out that significant variations in the overall C+N+O, heavy elements and s-process are associated with the horizontal position of the stars in the chromosome maps as well i.e. type II clusters in \cite{Milone17}.}

With this knowledge in hand, it is possible to identify a population in the chromosome maps of several clusters that would have significant changes in He at a constant N abundance consistent with field stars \citep[cf.][]{Milone17}. From now on we will refer to this population as extended P1 (P1 being the population in globular clusters with N abundance akin to field stars of similar [Fe/H]). The extended P1 cannot be accounted for by photometric uncertainties alone, for example, the difference between the most extreme stars in population A and B is $>15$ times the size of the mean photometric uncertainties for these stars (cf. Fig.~\ref{fig:cmd_cmap}), suggesting that it's presence reflects genuine He abundance variations among these stars \citep[cf.][]{Lardo18}. If this interpretation were to be correct, this population would have a deep impact on our understanding of origin of the abundance variations characteristic of globular clusters, as it would be a chemical pattern very difficult to account for with the current models \citep[cf.][more on this in \S~\ref{sec:int} and~\ref{sec:conc}]{Lardo18}.

The broad consensus among these models/scenarios is that nucleosynthesis via the CNO-cycle is responsible for producing He variations and changes in other elements like C, N, O, Na, Mg and Al that have been inferred though photometry and directly measured in spectroscopic campaigns. With the main difference between models being the site where these reactions take place, which include in principle: the core of massive main sequence stars;  the H-burning shell in stars at the tip of the RGB and RGB clump; as well as in the bottom of the convective envelope in (super)AGB stars \citep[cf.][and references therein]{Charbonnel16}.

During the CNO-cycle, nuclear fusion turns H into He with the help of C, N and O as catalysts (where C and O turn into N, and so changing their relative abundances). If the temperature increases, the NeNa-chain activates with a net result of depleting O while enhancing Na. An additional increase in temperature triggers the MgAl-chain which ends up  producing Al and depleting Mg \citep[cf.][and references therein]{CS13,Charbonnel16}. These abundance patterns qualitatively agree with the observed abundance variations among globular cluster stars\footnote{\label{note1}Although there are some caveats with this picture e.g. \cite{BCZS15,Prantzos17,Carretta18}.}, as a consequence the origin of the latter was linked to these reactions \citep[cf.][]{BL18}.

\cite{Milone15} identified five sub-populations in the chromosome map of NGC 2808, two of them (photometric groups A and B) among the extended P1 (cf.  Fig. \ref{fig:cmd_cmap}). The abundances of several elements have been determined for stars belonging to group B \citep[e.g.][]{Milone15,Carretta18}. These have been found to be consistent with the values of field stars of similar [Fe/H] however, the abundance of N have not been investigated\footnote{\cite{Carretta18} reported the relative CN abundance of 89 stars, a rough proxy for N. They found that the CN abundance of group B stars was lower than any of the other groups (excluding group A which was not represented in their sample).}. However, the chemical composition of stars from photometric group A have not been measured to date, although they are key to constraining what kind of chemical differences are present within population P1 in the chromosome map.

In this work, we present spectroscopic abundances (including [N/Fe]) of the extended P1 of NGC 2808, a cluster with one of the largest extended P1 in \cite{Milone17}, sampling for the first time stars from groups A and B with the aim of testing the predicted homogeneity in the abundances of CNO-cycle products (e.g. C, N and O) proposed by different interpretations of the chromosome map.

\section{Data}

The present work is centred in the analysis of high resolution spectra of six stars, all of them members of NGC 2808 according to proper motions and radial velocities (cf. Table~\ref{tab:1}). These stars were chosen to occupy very similar regions in the optical CMD in order to minimise effects arising from differences in stellar parameters, however, they stretch along most of P1 extent in the chromosome map, sampling both population A and B defined in \cite{Milone15} (cf. Fig.~\ref{fig:cmd_cmap}). The models of \cite{Lardo18} suggest that the difference in the He mass fraction between population B and A is of $\Delta \mbox{Y}\sim 0.06$ (assuming the standard $\mbox{Y}=0.246$ for population A). The same models also suggest that [N/Fe] should be the same in both populations.

The spectra were taken with MIKE at the Magellan-Clay Telescope \citep{Bernstein03} during the nights of January 15-16, 2018, using the $0.7\arcsec\times5\arcsec$ slit, providing a spectral resolution of $\sim 36,000$ in the red arm. The data was reduced with the {\tt CarPy} version of the pipeline \citep{Kelson03} and corrected for tellurics using the {\tt telluric} task in {\tt IRAF}. In addition to the scientific targets, two spectrophotometric standards (GD 71 and LT 3218) were observed with the $2\arcsec\times5\arcsec$ slit to allow a relative flux calibration of the science targets.

The photometry used throughout the paper is from the catalogues published by \cite{Soto17}, and the data for the chromosome maps are from \cite{Lardo18}. Finally, we note that all our stars are consistent with the radial velocities previously reported for cluster members \citep[systemic RV $=101.4\pm 1$ km/s and central radial velocity dispersion $\sigma=18.8\pm4.0$ km/s, cf.][]{Lardo15kin} and proper motions \citep[i.e. all targets with a 0.0 displacement in pixels after $\Delta$Epoch = 7.5 yr, cf. D\_x and D\_y values in the][ catalogs]{Soto17}. This implies that these stars are unlikely members of the field.

\section{Analysis}

The atmospheric parameters were determined from the $VI$ photometry of our targets. Effective temperature T$_{\mbox{eff}}$ values were obtained from ($V-I_{C}$) colours using the \citet{alonso99,alonso01} calibration and adopting a colour excess $\mbox{E}(B-V)=0.22$. Initial surface gravity ($\log$(g)) values were derived using the standard formula:
$$ \log{\left(\rm \frac{g}{g_{\odot}}\right)} = 0.4({\rm M_{V}} + {\rm BC} - {\rm M_{bol,\odot}}) + 4\log \left(\frac{{\rm T_{eff}}}{{\rm T_{eff,\odot}}}\right) +\log \left(\frac{{\rm M}}{{\rm M_{\odot}}}\right) $$
assuming M$_{{\rm bol,\odot}}$ = 4.75, T$_{\mbox{eff}, \odot}$ = 5777 K, and $\log${\rm (g)}$_{\odot}$ = 4.44 for the Sun. We adopted a stellar mass of 0.8 M$\odot$ and  calculated the bolometric corrections (BCs) using the calibration given in \citet{alonso99}. Finally, initial microturbulent velocities ($v_t$) were estimated to be 1.5 km/s for all stars in our sample. From that initial guess, the final $v_t$ values were obtained by erasing any trend between Fe{\sc i} and the logarithm of the reduced equivalent widths ($\log$(RW) = $\log$(EW/$\lambda$)). We note that the spectroscopic parameters are consistent, within the uncertainties, with the photometric ones and the use of the former do not change the results discussed here.

We determined the chemical abundances for all the studied elements (but C and N) through the measurement of the equivalent widths (EWs) of atomic transition lines and employing the package {\tt GALA} \citep{Mucciarelli13} based on the {\tt width9} code by R. L. Kurucz. Model atmospheres were calculated with the {\tt ATLAS9} code
assuming local thermodynamic equilibrium (LTE) and one-dimensional, plane-parallel geometry; starting from the alpha enhanced grid of models available in F. Castelli's website \citep{castelli03}. For all the models we adopted an input metallicity of [A/H]=-1.15 dex, according to the value given in \citet{Carretta15}.

For C and N, we measured the G-band feature at 4300~\AA~and the CN $UV$ feature at $\sim$3880~\AA; respectively.
Abundances were measured through a $\chi^2$-minimization between the observed spectrum and a grid of synthetic spectra calculated at different abundances. The synthetic spectra have been computed by means of {\tt SYNTHE} code developed by Kurucz, including the entire Kurucz/Castelli line-list (both for atomic and molecular transitions) and adopting the same model atmospheres used to derive the abundances from EWs. 

Corrections for departures from the LTE approximation are applied only for the Na lines, by interpolating on the grid of corrections calculated by \citet{Lind11}\footnote{from the INSPECT database, version 1.0 \url{http://inspect.coolstars19.com}}. Reference solar abundances are from \citet{GS98}.

The uncertainties in the chemical abundances were obtained as discussed in e.g. \citet{mucciarelli17}. For those elements for which abundances have been measured using EWs, the random errors are computed as the dispersion of the mean normalized to the root mean square of the number of used lines. For the elements measured via spectral synthesis, the random error has been estimated using Montecarlo simulations, adding Poissonian noise into the best-fit spectrum in order to reproduced the measured signal-to-noise ratio and creating for each star a sample of 100 noisy synthetic spectra. These spectra have been re-analysed with the same procedure used for the observed spectra and the $1\sigma$ dispersion of the abundance distribution taken as uncertainty. Uncertainties in chemical abundances due to the adopted model atmospheres were estimated by varying the stellar parameters, one at a time, in steps equal to the uncertainties in T$_{\mbox{eff}}$, $\log$(g) and $v_t$ respectively (cf. Table~\ref{tab:1}).

\begin{table*}
\begin{tiny}

\caption{Radial velocities, stellar parameters and abundances of our targets.} \label{tab:1}
\begin{tabular}{cccccccccccccccccc}
Star     & RV & T$_{\mbox{eff}}$ & log(g) & $v_t$ & [C/Fe]  & [N/Fe]  & [O/Fe] & [Na/Fe] (NLTE) & [Mg/Fe]  & [Al/Fe] & [Si/Fe] & [Fe/H] \\
ID*	& (km s$^{-1}$) & (K) & (cm s$^{-2}$) & (km s$^{-1}$) & (dex) & (dex) & (dex) & (dex) & (dex) & (dex) & (dex) & (dex) \\
\hline
55974$^{\mbox{a}}$  & $95.90 \pm 0.89$ & 4809 & 2.15 & 1.30 & $0.05 \pm 0.16 $& $0.08 \pm 0.17$ & $0.37 \pm 0.16$ &  $-0.33 \pm 0.05$ & $0.18 \pm 0.06$ & $-0.04 \pm 0.09$ & $0.18 \pm 0.09$ & $-1.10 \pm 0.09$ \\
57615$^{\mbox{a}}$   & $81.08 \pm 1.08$ & 4796 & 2.08 & 1.30 & $-0.20 \pm 0.13$ & $0.05 \pm 0.17$ & $0.39 \pm 0.13$ &  $-0.51 \pm 0.08$ & $0.11 \pm 0.06$ & $-0.06 \pm 0.11$ & $0.17 \pm 0.10$ & $-0.87 \pm 0.09$ \\
126605$^{\mbox{b}}$  & $86.30 \pm 1.10$ & 4869 & 2.14 & 1.10 & $-0.22 \pm 0.13$ & $0.13 \pm 0.19$ & $0.40 \pm 0.17$ &  $-0.35 \pm 0.05$ & $0.10 \pm 0.06$ & $-0.02 \pm 0.07$ & $0.08 \pm 0.08$ & $-1.08 \pm 0.09$ \\
187128$^{\mbox{b}}$  & $105.87 \pm 1.46$ & 4776 & 2.01 & 1.70 & $-0.03 \pm 0.14$ & $0.40 \pm 0.19$ & $0.35 \pm 0.12$ &  $-0.17 \pm 0.04$ & 	-- **		    & $0.04 \pm 0.08$ & $0.15 \pm 0.07$ & $-0.92 \pm 0.07$ \\
276739$^{\mbox{b}}$  & $118.24 \pm 1.40$ & 4851 & 2.15 & 1.00 & $0.00 \pm 0.16$ & $-0.04 \pm 0.17$ & $0.45 \pm 0.14$ &  $-0.36 \pm 0.06$ & $0.07 \pm 0.06$ & $-0.11 \pm 0.09$ & $0.10 \pm 0.09$ & $-1.04 \pm 0.09$ \\
298744$^{\mbox{a}}$  & $97.76 \pm 1.13$ & 4851 & 2.18 & 1.40 & $0.00 \pm 0.12$ & $0.21 \pm 0.17$ & $0.49 \pm 0.17$ &  $-0.41 \pm 0.04$ & $0.18 \pm 0.04$ & $-0.07 \pm 0.07$ & $0.09 \pm 0.08$ & $-0.96 \pm 0.08$ \\
\hline
34250$^{\mbox{d}}$  & $110.37  \pm 1.86$  & 4490 & 1.50 & 1.57 & -- 				 & -- 			 & -- 				& $0.04 \pm 0.05$ & $-0.08 \pm 0.04$ & --  & $0.18 \pm 0.07$ & $-1.05 \pm 0.06$ \\
127544$^{\mbox{d}}$  & $111.88  \pm 1.38$ & 4437 & 1.41 & 1.38 & -- 				 & -- 			 & -- 				& $0.07 \pm 0.05$ & $0.00 \pm 0.04$ & --  & $0.24 \pm 0.06$ & $-1.09 \pm 0.06$ \\
184806$^{\mbox{c}}$  &  $112.13 \pm 2.35$ & 4725 & 1.87 & 1.04 & -- 				 & -- 			 & -- 				& $-0.20 \pm 0.04$ & $0.14 \pm 0.04$ & --  & $0.07 \pm 0.07$ & $-0.97 \pm 0.06$ \\
257773$^{\mbox{d}}$  & $112.37  \pm 1.10$   & 4486 & 1.49 & 1.49 & -- 				 & -- 			 & $-0.04\pm0.13$	& $-0.01 \pm 0.05$ & $0.06 \pm 0.15$ & -- & $0.17 \pm 0.09$ & $-1.06 \pm 0.08$ \\
\end{tabular}

Typical errors in T$_{\mbox{eff}}$, log(g) and $v_t$ are $\pm75$K,  $\pm 0.2$ cm s$^{-2}$ and  $\pm 0.2$ km s$^{-1}$ respectively.
*The letters next to the ID of the stars represent the photometric group to which they are most likely associated.
 **Mg for this star is in the flat par of the curve of growth, hindering a reliable estimate.

\end{tiny}
\end{table*}

\begin{figure}
\centering
    \includegraphics[width=\columnwidth]{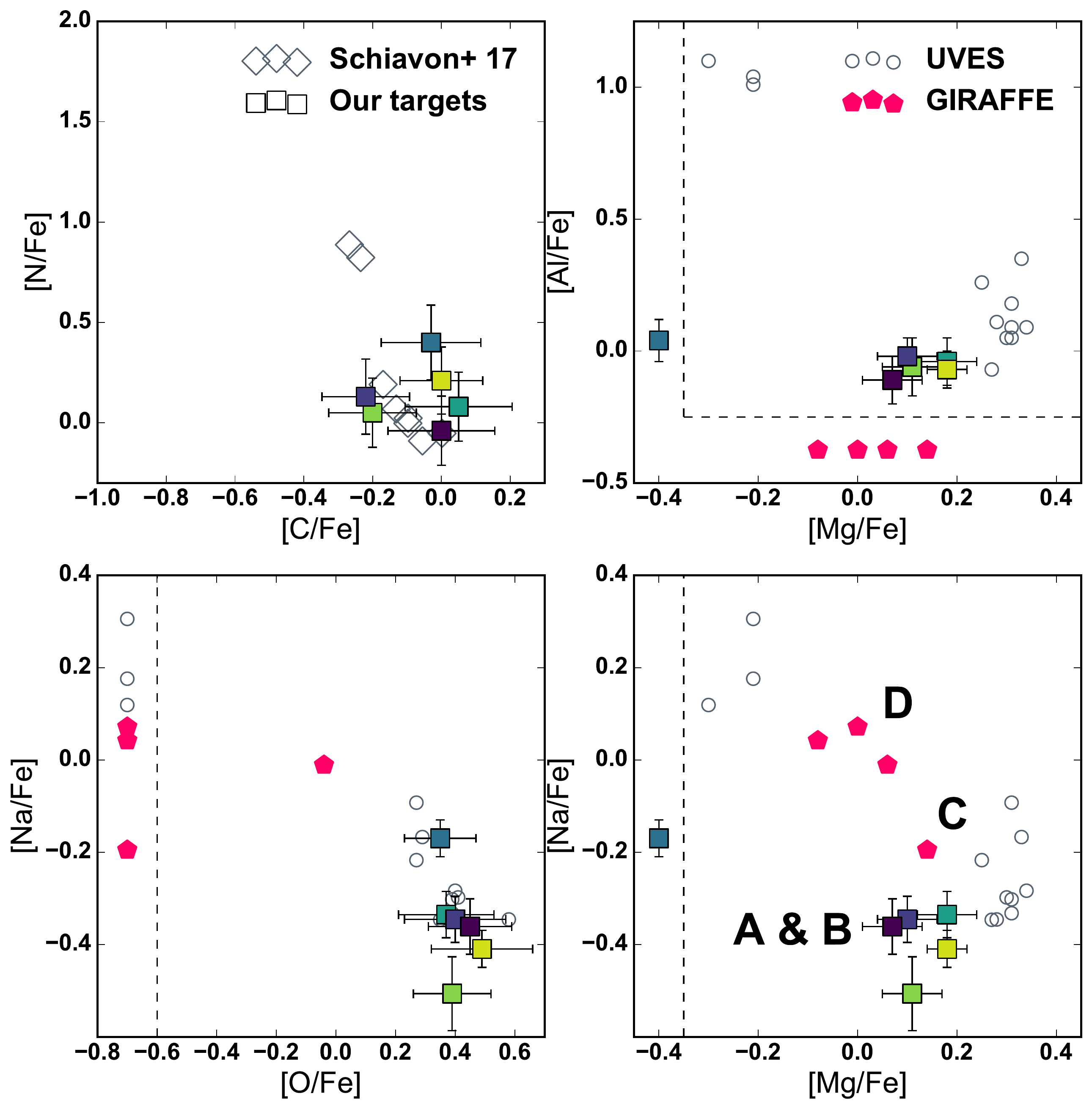}
    \caption{Comparison of abundances our targets stars (square symbols) against reference data (other symbols). The colour of the square symbols represent the $\Delta$(F275W-F814W) value of the P1 stars in our sample cf. Fig.~\ref{fig:cmd_cmap}. \textbf{Upper left:} N-C abundances of M 107 stars (similar metallicity than NGC 2808) as reference. The other panes display as a reference abundances from archival UVES and GIRAFFE data of NGC 2808. \textbf{Upper right:} Al-Mg plane;  \textbf{Lower left (right):} Na-O (Na-Mg) plane. The pink symbols are P2 stars in common with \protect\cite{Lardo18} chromosome map. Stars without [element/Fe] messurements were assigned arbitrary values smaller than the respective dashed line). The letters in the Na-Mg plane denote their positions in the chromosome map from Fig.~\ref{fig:cmd_cmap}. In none of the cases we find evidence of significant abundance variations among our targets.}
    \label{fig:abu}
\end{figure}

Fig. \ref{fig:abu} shows the abundance of the main elements known to change the most from star-to-star in GCs (the main products of hot H-burning). The squares in the figure denote stars in our sample while the diamonds, circles and pentagons reference literature and archival data. For C and N we use the abundances of M 107 (similar [Fe/H] than NGC 2808) from \cite{Schiavon17} as reference. For Na, O, Al and Mg we use as a reference abundances derived from UVES- and GIRAFFE-VLT archival data of NGC 2808 stars (programme ID: 072.D-0507). This archival data was analysed in the same way as our extended P1 targets.

\section{Interpretation}
\label{sec:int}

In agreement with previous studies \citep[e.g.][]{Milone15,Carretta18}, Fig. \ref{fig:abu} shows that the abundances of photometric group B, is consistent with abundances of field stars with similar [Fe/H], while stars from photometric groups C and D show abundance variations with respect of field stars of similar [Fe/H] (i.e. high Na with low O and Mg). Furthermore this figure also shows that the stars in our sample from photometric groups A and B in the chromosome maps are homogeneous in the abundances C, N, O, Na, Mg and Al (i.e. both with field like composition).

\begin{figure*}
 \centering
     \includegraphics[width=\linewidth]{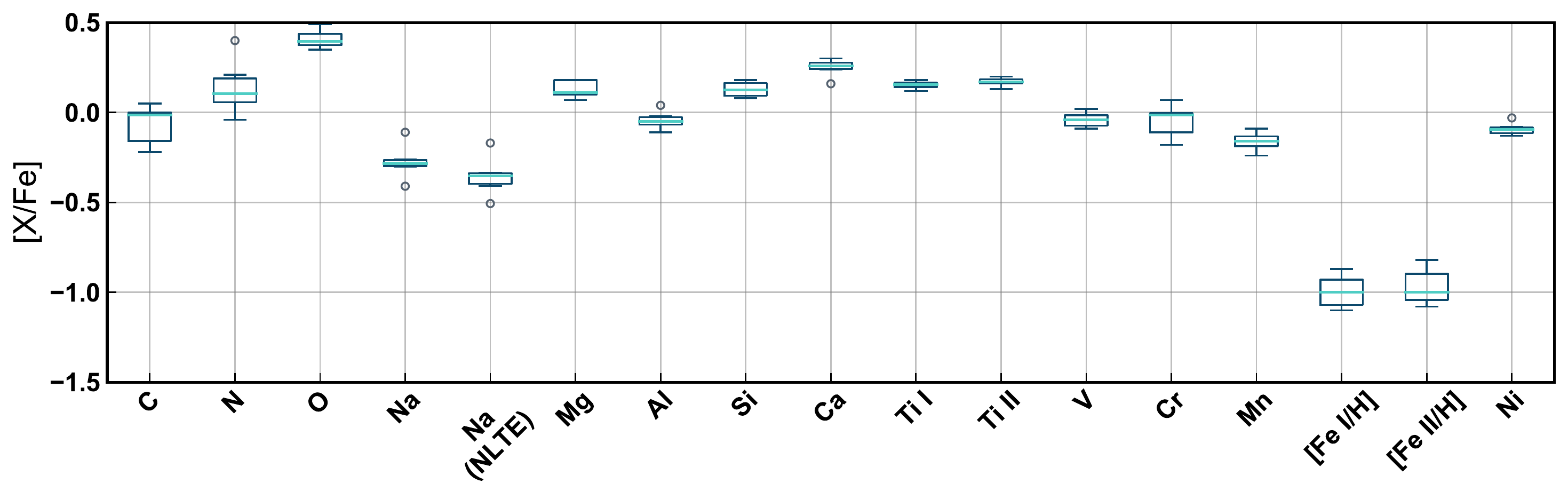}
     \caption{Tukey boxplot of the measured abundances of our targets. For each element, the blue box extends from the lower to upper quartile of the distribution of abundances. The cyan line in each box represents the median of the distribution, while the ``whiskers'' extend to 1.5 times the value of the interquartile range, in either side of the box. Data points that lie outside the whiskers are represented with circles. From this figure one can conclude that there is no evidence for significant abundance spread among the stars of photometric groups A and B in our sample in any of these elements.}     \label{fig:boxplot}
 \end{figure*}

Changes in the abundance of O, Mg and Si have been suggested to influence where stars lie along the $\Delta$(F275W-F814W) in the chromosome map, namely: changes in O could change the photometry of F275W as a relatively strong OH band falls within the passband of the filter; while changes in Mg and Si could affect the T$_{\mbox{eff}}$ of metal poor stars \citep[cf.][and references therein]{Lardo18}. However, these effects do not seem to be responsible for the $\Delta$(F275W-F814W) positions of the stars in our sample, they all appear to be homogeneous in the abundances of these elements and the the largest T$_{\mbox{eff}}$ difference between our stars is only $\sim90$ K, cf. Table~\ref{tab:1} and Fig.~\ref{fig:boxplot}. No evidence for a significant heavy element variations is found in our sample either, cf. Fig.~\ref{fig:boxplot}. This suggest that the mechanism responsible for the horizontal extent of our stars in the chromosome map, is of a different nature than the one affecting the type II clusters described in \cite{Milone17}.

The absence of a significant variation in the abundances of these CNO-cycle products among our sample of P1 stars of different photometric groups suggest that a different mechanism is responsible for these photometric variations. This would be in stark contrast with the current paradigm in which the photometric sub-populations found among globular cluster stars are the consequence of abundance variations produced by CNO-cycle nucleosynthesis \citep[cf. reviews by][and references therein]{Gratton04,Gratton:2012p2005,Charbonnel16,BL18}.

An interpretation that would fit the current evidence would be the one proposed by \cite{Lardo18} and \cite{Milone18}. This is, that changes in the temperature among the extended P1 stars are responsible for this photometric signature in the chromosome maps. In this hypothesis, these temperature changes are a consequence of different He mass fractions among the extended P1 stars. However, these works explicitly state, that this is premised upon the extended P1 stars being homogeneous in C, N and O. At the time of writing, there is no direct evidence of He variations among the extended P1 stars (e.g. spectroscopic studies of the strength of He lines in P1 stars of different temperature), so they remain conjectures.

\subsection{Implications of the possible He variations among extended P1 stars}

Under the working hypothesis that the results of synthetic spectral models by \cite{Lardo18} and \cite{Milone18} are correct, and that changes in $\Delta$(F275W-F814W) reflects the consequences of He variations, we can identify a possible nucleosynthetic channel for its production.

As mentioned above He is the primary product of H-burning, and the current evidence would indicate that at least two different channels are in play during the production of He rich material found in globular cluster stars: one producing the well studied He, C, N, O, Na, Mg and Al anticorrelations (akin to the products of nucleosynthesis through the hot CNO-cycle, that would describe P2 stars) and the other producing He variations without any corresponding change to other light elements (similar to the ashes of the proton-proton chain, which would be more appropriate for the evidence from our sample of P1 stars). We refer the interested reader to \S 8 of \cite{Milone18} for a brief discussion about some sources for material processed by p-p chain reactions that could produce the suggested He variations among P1 stars and some their associated caveats.

Future studies with larger sample sizes are needed to explore possible correlations between $\Delta$(F275W-F814W) and properties like [Fe/H] or T$_{\mbox{eff}}$ that could influence the relative position of the stars along the extended P1 population of the chromosome map in reliable manner.

\section{Summary and conclusions}
\label{sec:conc}

Previous studies have been able to isolate distinct sub-populations of globular cluster stars with different light element abundances in several photometric planes \citep[e.g.][]{Milone15,Carretta18}. The pseudo colour-colour plot so called ``chromosome map'' (a (F275W-F336W) - (F336W-F438W) vs. F275W-F814W plot normalised in a convenient way) is among the most popular due to its efficacy. By modelling stellar atmospheres different groups have reached the conclusion that the vertical axis is mostly sensitive to variations in the N abundance of stars, while the horizontal axis is mostly sensitive to relative temperature changes of stars at a given magnitude, which have been speculated to be the manifestations of differences in the He mass fraction among these stars \citep[cf.][]{Milone15,Lardo18}.

According to this interpretation, the chromosome map of some clusters reveal the presence of a new sub-populations of stars (the extended P1, composed of stars from group A and B in Fig. \ref{fig:cmd_cmap}) which does not obey the longstanding paradigm which proposes that these populations are the consequence of abundance patterns produced by hot H-burning through the CNO-cycle. These interpretations presented independently in the works of \cite{Lardo18} and \cite{Milone18} suggest that C, N and O (elements processed via this nucleosynthetic channel) should remain constant among the stars of this population according to their models.

So far, only abundances of some elements (excluding N) of photometric group B stars have been reported in the literature. In this work we target stars from photometric groups A and B of NGC 2808 (a cluster with one of the largest extended P1) and analyse their abundances to test proposed interpretations of the chromosome map.

From our analysis we find that all these stars (from groups A and B) are homogeneous in C, N, O, Na, Mg and Al abundances --all products of the CNO-cycle, which are known to change significantly from star-to-star in other photometric groups-- and other heavier elements, and are consistent with abundances of field stars with similar [Fe/H]. This represents a problem in the classical framework for the origin of multiple stellar populations in globular clusters, as this chemical signature is not expected and cannot be accounted in the current paradigm.

\emph{At the moment, the presence of He variations among these stars remains speculative. However, the challenge to the old paradigm presented here does not rely on the confirmation of He spreads among these stars, but in the lack of variations in the other CNO-cycle products in stars of different photometric groups.} So, the contention for a new paradigm for GC formation and their multiple stellar populations can include now these results to the list of discrepant evidence \citep[for other caveats see e.g.][]{cz15,Prantzos17,BL18,Carretta18}.

Having said that, the confirmation of the presence or absence of He variations among the extended P1 stars could provide important clues to its origin. For example, if He variations are present in stars the stars with constant CNO-cycle products, one could narrow down the physical mechanisms that can produce such He enhancement to something that resembles p-p chain nucleosynthesis.

Although proper spectroscopic He studies are challenging, specially in cool stars, they are not impossible \citep[e.g.][]{Dupree11,Pasquini11}. The results presented in this letter create a strong case for these kind of observations. 

\section*{Acknowledgements}

We would like to thank A. Dotter, M. Salaris, C. Johnson, C. Conroy and N. Bastian for insightful discussions and comments on an earlier version of this manuscript. Support for this work was provided by NASA through Hubble Fellowship grant HST-HF2-51387.001-A awarded by the Space Telescope Science Institute, which is operated by the Association of Universities for Research in Astronomy, Inc., for NASA, under contract NAS5-26555. CL acknowledges financial support from the Swiss National Science Foundation (Ambizione grant PZ00P2\_168065).




\bibliographystyle{mnras}
\bibliography{eP1_chemistry}




%
%


\bsp	
\label{lastpage}
\end{document}